\documentclass[%
 reprint,
 amsmath,amssymb,
 aps,
]{revtex4-2}

\usepackage{graphicx}
\usepackage{dcolumn}
\usepackage{bm}
\usepackage{amsmath}
\usepackage{bbm}
\usepackage{rotating}
\usepackage{float}
\usepackage{xcolor}
\usepackage{subcaption}
\usepackage{verbatim}

\usepackage{hyperref}

\newcommand{\bra}[1]{\ensuremath{\left\langle#1\right|}}
\newcommand{\ket}[1]{\ensuremath{\left|#1\right\rangle}}

\begin{document}

\preprint{APS/123-QED}

\title{Scalable architecture for measurement-induced squeezed light interferometers}

\author{Abhinav Verma}
\email{abhve@dtu.dk}
\author{Jacob Hastrup}
\altaffiliation{Present address: Center for Quantum Devices and NNF Quantum Computing Programme, Niels Bohr Institute, University of Copenhagen, Denmark}
\author{Jonas S. Neergaard-Nielsen}%
\author{Ulrik L. Andersen}%
\email{ulrik.andersen@fysik.dtu.dk}
\affiliation{Center for Macroscopic Quantum States (bigQ), Department of Physics, Technical University of Denmark, Fysikvej, 2800 Kongens Lyngby, Denmark}

\date{\today}

\begin{abstract}
Scalable interferometers lie at the heart of photonic quantum technologies, but their expansion has been fundamentally limited by optical losses that grow with circuit depth. Here, we introduce and experimentally demonstrate a measurement-induced architecture for multimode squeezed-light interferometers that overcomes this barrier. By shifting complexity from deep optical networks to programmable homodyne measurements, we realize effective transformations within a shallow, low-loss platform. We validate the principle with a six-mode device and extend it to a 400-mode interferometer, marking a leap in scale beyond conventional designs. Crucially, this strategy not only enables scalable squeezed light interferometry but also provides a powerful route to the generation of large-scale entangled states — a key requirement for quantum computing, simulation, and communication. Our results establish measurement-induced circuits as a practical pathway toward noisy intermediate-scale quantum (NISQ) applications, and future demonstrations of quantum advantage. 
\end{abstract}

\maketitle


\section{Introduction}
Multimode squeezed light interferometers, that is, multimode interferometers effecting a linear transformation on multiple input squeezed states, have become a cornerstone in the field of continuous variable photonic quantum information processing. They support a broad array of applications that extend across quantum computing, sensing, and communications. Such interferometer systems are not only instrumental in generating large-scale entangled graph states \cite{cluster19,GG1}, but also underpin advanced architectures for quantum neural networks and machine learning algorithms \cite{QNN1}. They form the core of Gaussian Boson Sampling \cite{GBS1,GBS2,Thekkadath2022,USTC20,Xanadu21, JZ4}, enabling both quantum advantage demonstrations and quantum simulation \cite{Paesani2019,Zhu2024}. Beyond computing, they provide essential tools for distributed quantum sensing \cite{DQS1,DQS2}, and the preparation of complex non-Gaussian resource states \cite{Su2019-ConversionGaussianStates,Eaton2022-MeasurementbasedGenerationPreservationa,Takase2023-GottesmanKitaevPreskillQubitSynthesizer,Larsen2025-IntegratedPhotonicSource}. Looking ahead, multimode squeezed light interferometers will be central to realizing the long-term vision of universal and fault-tolerant photonic quantum computing. In particular, when integrated with highly non-classical states such as Gottesman–Kitaev–Preskill (GKP) or cubic phase states, they offer the architectural backbone required to achieve fault tolerance and scalability. 


Numerous realizations of multimode squeezed light interferometers have been achieved, employing various technological avenues such as free-space optics \cite{JZ4,DQS2,USTC20,100Teleportation}, photonic integrated circuits \cite{Paesani2019,Larsen2025-IntegratedPhotonicSource}, fiber optics \cite{Larsen21,Xanadu21,USTC20}, or even spectral mode mixing \cite{Chen2014,Roslund2014,Cai2017}. In some cases, the interferometer circuit is fully programmable while in others, the programmability is highly limited or even reduced to a single configuration. Notably, the realizations using photonic integrated circuits are often fully programmable, enabling an arbitrary unitary Gaussian transformation. However, despite the promise shown by scalable on-chip interferometers with full programmability, the ambitious goal of scaling up to hundreds of modes is severely constrained by the optical losses that are significantly higher than in free-space or fibre optics. Furthermore, as the depth of the circuit increases the losses incurred also significantly increase.
Thus, it is unlikely that fully programmable interferometers can be scaled to a large mode number.

\begin{figure*}
\includegraphics[width=\textwidth]{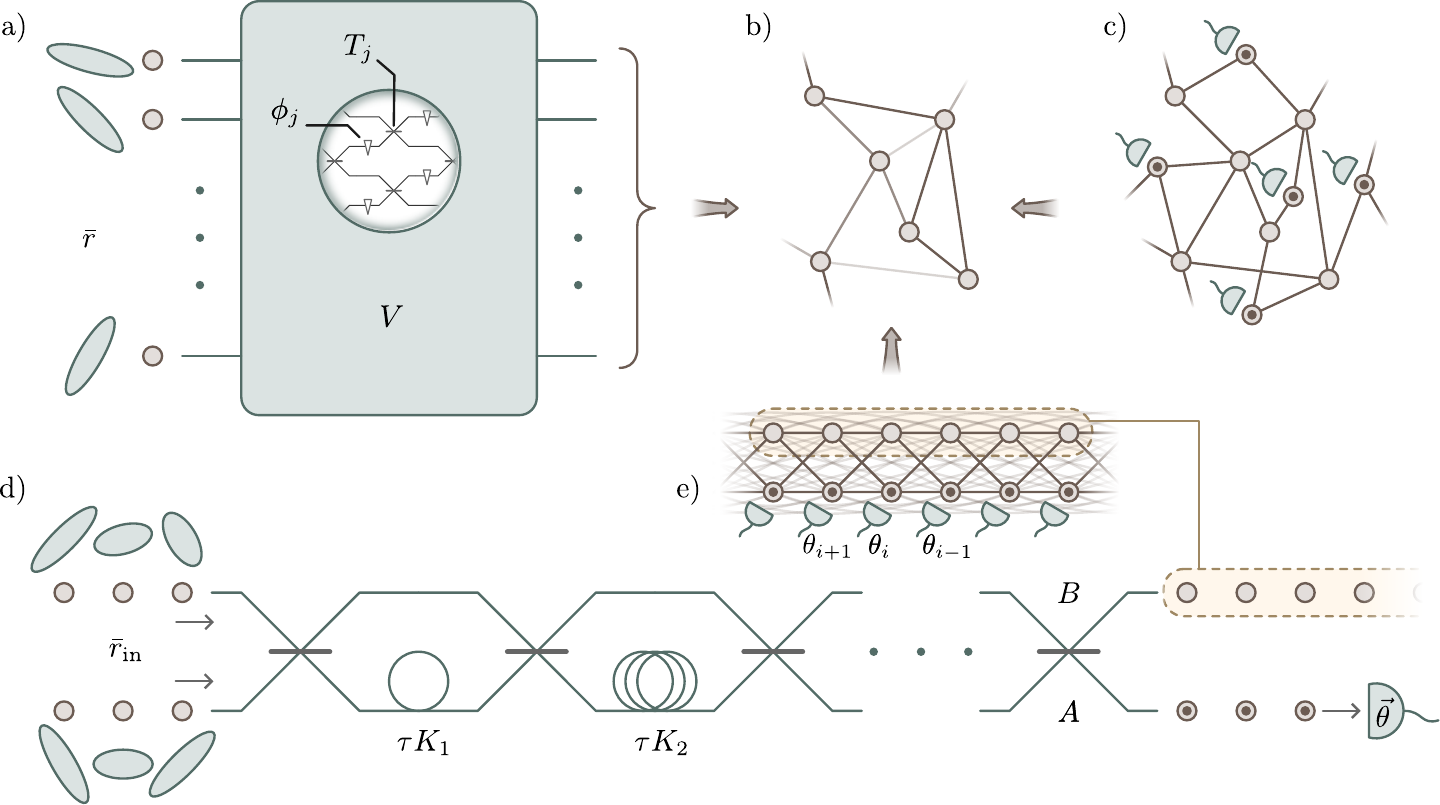}
\caption{\label{fig:cluster}
 Concept of measurement-induced squeezed light interferometers.
 a) Injecting $N$ single-mode squeezed states of light into a multimode interferometer with beam-splitters $\vec T$ and phase shifters $\vec\phi$, that implements an arbitrary linear transformation described by an $N\times N$ unitary matrix $V$, results in an entangled Gaussian state that can be described by a graph, b), where nodes and edges represent modes and correlations between them, respectively.
 c) A different way of obtaining the same graph state is to start from a larger graph state of $M$ modes and perform homodyne measurements on $M-N$ modes. The original correlations and the homodyne basis settings determine the resulting graph.
 d) The particular scheme we study in this paper consists of generating the initial graph state by injecting a sequence of squeezed states at time intervals $\tau$ into a two-mode, shallow-depth interferometer made up of unbalanced Mach-Zehnder interferometers with delay lengths of $\tau K_i$, for different integers $K_i$. This results in a multimode temporally encoded cluster state \cite{Yokoyama13,cluster19}, as illustrated by the graph in e). By partially measuring out this cluster state with variable homodyne basis settings $\vec\theta$, a reduced state corresponding to a graph as in b) is obtained. Note that the in-principle endless cluster state can be reduced to a finite size by $x$-basis measurements that remove correlations \cite{Menicucci2011}. In our experiments, the input squeezing $\vec r_\text{in}$ is constant and $K_1=1$, $K_2=8$.
 }
\end{figure*}

While full programmability is preferred, it is not a mandatory prerequisite for efficiently exploring the solution space relevant to specific problems across all quantum algorithmic applications. For example, hybrid quantum-classical algorithms do not necessarily require the quantum devices to possess full universality. E.g., in the case of the variational quantum eigensolver \cite{Peruzzo2014-VariationalEigenvalueSolver}, employed for determining the ground state energy of molecular systems, the parametrized circuit undergoes continuous adjustments based on the feedback from measurement outcomes until it reaches convergence. Provided that the circuit configuration resides within the viable solution space, there is no need for comprehensive programmability of the parameters, demonstrating a more flexible approach to quantum problem-solving.

In this article, we introduce and experimentally validate a non-universal, but highly scalable and low-loss multimode squeezed light interferometer that hinges on the strategic use of measurement-induced operations to enable partial programmability. Harnessing a continuous-variable Gaussian  state entangled in multiple temporal modes combined with effective interactions induced by high-efficiency homodyne  measurements, we facilitate the implementation of an effective multimode squeezed light interferometer through an exceptionally shallow circuit architecture. 
We first demonstrate the basic principle by realizing a compact 6-mode squeezed light interferometer and demonstrating the tunability of such a system in producing a measurement-based graph state. Following this, we underscore the potential for scalability by successfully extending the interferometer to incorporate 400 squeezed modes. Our findings unlock the potential for large-scale implementations of NISQ algorithmic applications, including the promising domain of hybrid quantum-classical algorithms, which are now feasible with the utilization of shallow circuit designs. 

\section{Concept and theory}

A conventional programmable squeezed light interferometer, shown in Fig.~\ref{fig:cluster}a, is a multiport interferometer, described by an $N\times N$ unitary matrix $V$ and consisting of variable beamsplitters $\vec{T}$ and phase shifters $\vec\phi$, fed by squeezed vacuum input states $\hat{S}(\vec{r})\ket{0}^{\otimes N}$. 
This can generate arbitrary (pure and zero-mean) Gaussian multi-mode states $|\Phi_\text{out}\rangle=
\hat{U}_V(\vec T,\vec\phi) 
\hat{S}(\vec r) |0\rangle^{\otimes N}$, where $\hat{U}_V$ is the unitary operation of the interferometer $V$ on the $N$-mode Hilbert space. 
Mathematically, such pure states can also be represented as graphs \cite{Menicucci2011-GraphicalCalculusGaussian} where vertices correspond to optical modes and edges indicate correlations (Fig.~\ref{fig:cluster}b). While such an interferometer provides full programmability -- allowing the construction of any desired graph through control of beam-splitting ratios, phases and input Gaussian states -- the circuit depth increases proportionally with the number of modes, making it impractical to scale up. To address this limitation, we propose a fundamentally different approach: shifting the programmability from the interferometer itself to a measurement system. By first generating an initial graph state of $M$ modes and then applying programmable quadrature measurements to $M-N$ modes (see Fig.~\ref{fig:cluster}c), a tailored graph with $N$ vertices is created: The correlations between the remaining $N$ modes will change depending on their initial correlations with the $M-N$ measured modes and on the choice of quadrature basis.

A specific class of initial $M$-mode graph states can be efficiently generated using a scalable platform based on time-domain correlations \cite{Menicucci2011}. 
A conceptual schematic of this graph state generator is depicted in Fig.~\ref{fig:cluster}d. The setup employs two sources of single-mode squeezed light with squeezing parameters $\vec{r}_\text{in}$ directed through a cascaded set of interferometers with highly unbalanced arm lengths. The delays in the interferometers are integer multiples,  $K_i$, of the clock cycle $\tau$ of the squeezed light. 
The resulting states manifest correlations across two spatial output modes, A and B, as well as temporally among sequentially produced modes (Fig.~\ref{fig:cluster}e), with the complexity of these correlations increasing with the addition of more interferometers. As examples, by using a configuration of a single or two interferometers, it is possible to generate one-~\cite{Yokoyama13} and two-dimensional~\cite{cluster19,GG1} graph states with correlations suitable for continuous-variable quantum computing~\cite{Larsen21a,Larsen21b}.


In the next phase, the programmability of the setup is realized through the homodyne measurement process applied to a selected subset of modes. Initially, we consider the case where all temporal modes of spatial mode $A$ are measured, giving $N=M/2$. Alternative measurement patterns will be discussed later. 
The quadrature angles of the homodyne detections are parametrized by $\vec{\theta}=(\theta_1,\theta_2,\ldots, \theta_{M-N})$. These measurements influence the entanglement structure of the remaining modes in spatial mode $B$, effectively tailoring the time-domain entangled state into a new $N$-mode graph state,
%
%
\begin{equation}
|\Phi_\text{out}\rangle=
\hat{U}_V(\vec T_\text{eff},\vec\phi_\text{eff}) 
\hat{S}(\vec r_\text{eff}) |0\rangle^{\otimes N},
\end{equation}
where $\hat{U}_V(\vec T_\text{eff},\vec\phi_\text{eff})$ and $\hat{S}(\vec r_\text{eff})$ denote the effective, measurement-induced operations acting on the  $N$ modes. 
In reality, this state will also be subject to a phase-space displacement dependent on the random homodyne measurement outcomes; however, as described below, we disregard this in order to focus on the induced correlations.
The parameters $\vec T_\text{eff},\vec\phi_\text{eff}$, and $\vec r_\text{eff}$ depend on the fixed circuit configuration $\{K_i\}$, the programmable measurement angles $\vec\theta$, and the initial squeezing parameters $\vec r_\text{in}$.
This measurement-induced programmability enables control of the properties of the output state while maintaining the scalability and low-loss benefits of the fixed interferometer architecture. Although the long delay lines in this system introduce some loss, each optical path is traversed only once, ensuring that the total loss does not scale with the number of modes.

Thus, the resultant time-domain output of mode B mirrors the output of a traditional squeezed light interferometer, where correlations span different spatial outputs of a linear beam splitter plus phase shifter network supplied with squeezed light as in Fig.~\ref{fig:cluster}a. 
This output state is Gaussian and can therefore be described by its covariance matrix $\sigma$.
To identify the effective circuit parameters for a given measurement-induced state,
the covariance matrix is decomposed into a symplectic transformation $S$ of a thermal state represented by a diagonal matrix $D$ of its thermal eigenvalues (Williamson decomposition), $\sigma=\frac{1}{2}SDS^{T}$.
Next, the Bloch-Messiah-Euler decomposition dissects the symplectic transformation into initial squeezing transformations $\hat{S}(\vec r_\text{eff})$ (with diagonal matrix $S_q$) and a passive linear transformation given by an orthogonal and symplectic matrix $L$. The unitary matrix $V$ and a corresponding unitary operator $\hat{U}_V(\vec T_\text{eff}, \vec\phi_\text{eff})$ is extracted from $L$ (see Supplementary Material, sec. 3 \cite{supp}). 
Further, the (non-unique) effective circuit parameters $\vec{T}_\text{eff}$, $\vec{\phi}_\text{eff}$ can be extracted from $V$ \cite{Reck1994,Clements2016}.
However, as mentioned above, a by-product of the homodyne measurements used to tailor the resulting Gaussian states are displacements on the remaining modes which depend on the random measurement outcomes. These must be either actively corrected by measurement feed-forward or corrected for in post-processing. 
If not, the effect of the accumulation of these displacements through multiple realisations will show up in the thermal eigenvalue matrix $D$. In our experimental results, we treat these displacements in post-processing at the ensemble level by setting $D=\frac{1}{2}\mathbb{I}$, thereby effectively nulling the displacements
(see Supplementary Material, sec. 3 \cite{supp}).

\section{Expressibility}


Our scheme is not universal; it cannot reproduce an arbitrary $N$-mode squeezed light interferometer. Nevertheless, it covers a broad subset of transformations. This becomes clear when comparing the $O(N^2)$ scaling of the number of circuit parameters in a fully programmable interferometer with the linear scaling of the programmable measurement parameters $\vec\theta$. Despite this restriction, the approach remains expressive enough to realize a wide range of tunable multimode transformations within a scalable and low-loss platform.
To gauge the scheme's versatility, we turn to the concept of expressibility.
The conventional definition of expressibility \cite{PQC1} compares the distribution of fidelities between pairs of output states randomly sampled from the ensemble $\mathcal{P}$ of all parametrized circuits with the equivalent distribution of states from a Haar-random reference ensemble $\mathcal{H}$.
Formally, it can be quantified as the squared Hilbert-Schmidt norm $||\hat{A}_t||_\text{HS}^2$ of the deviation from a state $t$-design,

\begin{equation}
    \hat{A}_{t}=\int_{\mathcal{H}} (\ket{\psi}\!\bra{\psi})^{\otimes t} d\psi - \int_{\mathcal{P}} (\ket{\phi}\!\bra{\phi})^{\otimes t}d\phi \ .
\label{deviation}
\end{equation}
As shown in the Supplementary Material, sec. 2 \cite{supp}, this is given by
\begin{widetext}
\begin{equation}
    ||\hat{A}_{t}||_\text{HS}^{2} = \int_\mathcal{H}\!\int_\mathcal{H} |\langle\psi|\psi'\rangle|^{2t} d\psi d\psi' + \int_\mathcal{P}\!\int_\mathcal{P} |\langle\phi|\phi'\rangle|^{2t} d\phi d\phi' - 2 \int_\mathcal{P}\!\int_\mathcal{H} |\langle\psi|\phi\rangle|^{2t}d\psi d\phi \ . 
\label{expressibility}
\end{equation}
\end{widetext}
The first two terms are the $t^\text{th}$ moments of the intra-set fidelities over $\mathcal{H}$ and $\mathcal{P}$ and the last term is the $t^\text{th}$ moment of the inter-set fidelities. Intuitively, this expression can be seen to represent the difference between the distributions of the intra-set and inter-set fidelities in terms of their $t^\text{th}$ order moments.
In practice, these expressions are estimated through sampling of states from both ensembles.
The closer $||\hat{A}_t||_\text{HS}^2$ is to zero for each $t$, the better the given parametrized circuit can reproduce the states of the reference ensemble, i.e. the more expressible it is.
However, this is not immediately a useful definition for infinite-dimensional systems such as the one we consider here, where the Haar measure does not exist, making the integrals over $\mathcal{H}$ ill-defined. Another way to think of it is that the infinitely many possible states of arbitrary energy in $\mathcal{H}$ results in vanishingly few non-zero pairwise fidelities.
To get a practical measure of the expressibility of our circuits, we need to tweak the definition by compactifying the space of reference states: Instead of the non-existing uniform distribution over all possible states $|\psi\rangle$, we narrow the reference ensemble down to states that can be represented similarly to the parametrized states, i.e.\ as a linear transformation $\hat{U}_W$ of squeezed input states $\hat{S}(\vec{r}_\text{ref})\ket{0}$. Since $W$ belongs to the compact group of unitary matrices, \textsl{this} can be sampled uniformly, thereby providing the most general reference ensemble for the interferometer itself.
The squeezing values now need to be confined, and the most natural choice is to let them follow the same distribution as that of the measurement-induced squeezing values, $p(\vec{r}_\text{ref}) = p(\vec{r}_\text{eff})$.
This recipe for populating the reference ensemble $\mathcal{H}$ results in states of same average energy as the states of $\mathcal{P}$, but where the measurement-induced linear optical network $V$, which is the main object of our interest, can be compared to truly Haar-random reference networks $W$.

\begin{figure*}
    \includegraphics[width=\textwidth]{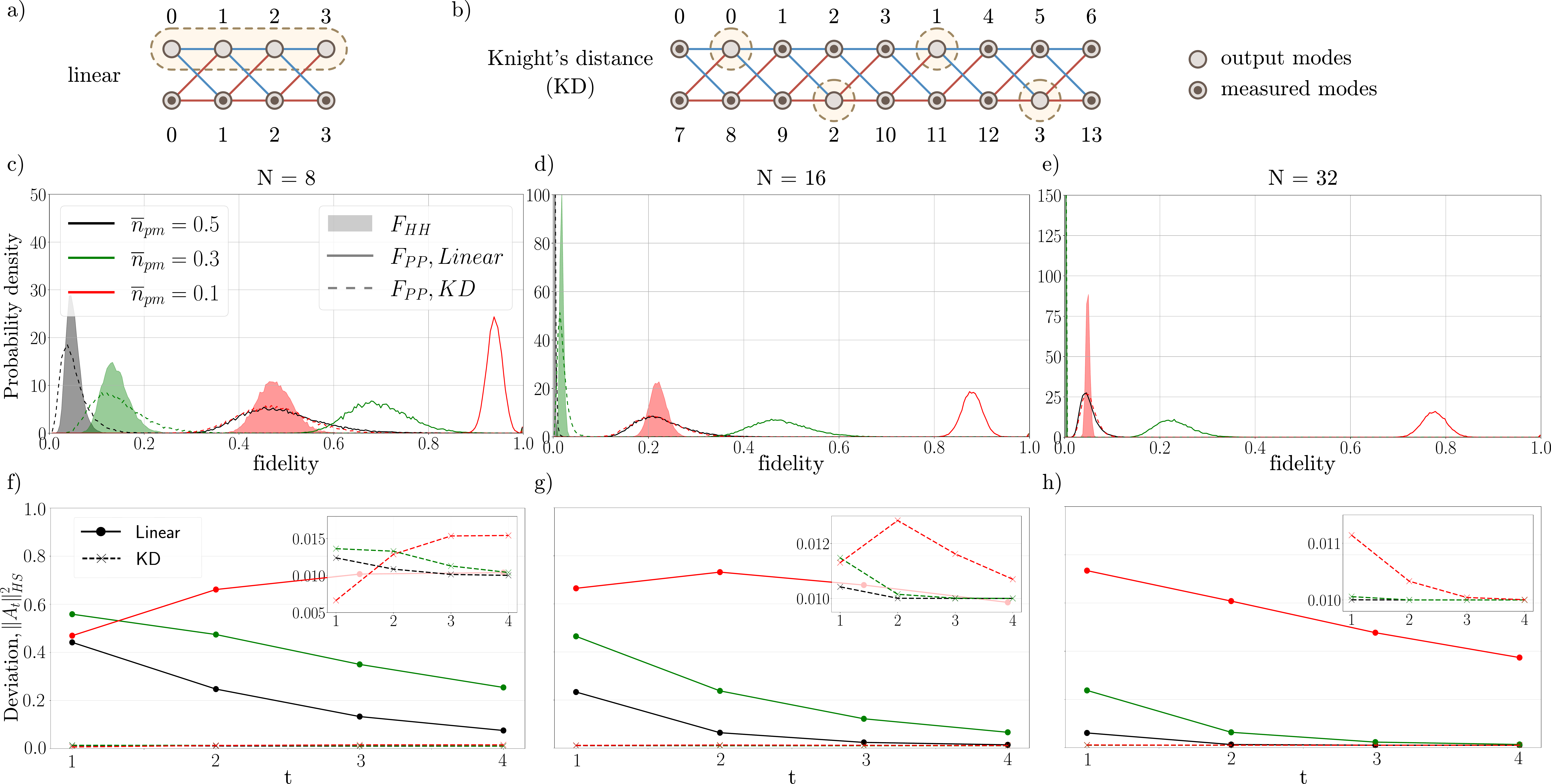}
        
     \hfill
     \caption{\label{fig:KDvLinF} 
    Simulated relative deviation of measurement-induced unitaries from reference unitaries. Comparison between two different measurement strategies on an initial 1D cluster state, as illustrated at the top: a straight-forward ``linear'' measurement pattern (a), vs.\ a ``Knight's distance'' (KD) pattern (b).
     The panels show, for interferometers of sizes $N=8$, $16$ and $32$, the distributions (c-e) and the deviations (f-h) of a set of parametrized states from a random ensemble for different values of $t$ and the average photon number per mode in the output state, $\overline{n}_{pm}$. The inset provides a zoomed-in view of the datapoints corresponding to the KD strategy. 
     Each dataset is generated by uniformly sampling the circuit parameters, with 1000 samples drawn per case, resulting in 1000 parametrized states and 1000 unitaries. The KD approach shows a clear advantage over the linear strategy, though this advantage gradually diminishes as $N$ increases.}
\end{figure*}

The expressibility of the parametrized circuits depends on several factors, which we explore in the following through simulation: 1) the architecture of the initial graph (i.e.\ the correlation complexity), 2) the number of modes, 3) the measurement strategy, and 4) the degree of input squeezing. 
Regarding 1), the complexity of the graph state correlations can be varied by increasing the number of cascaded interferometers in the setup as shown in Fig.~\ref{fig:cluster}. However, for this simulation we fix the initial graph as a 1D cluster state, that is, with only one delay line, $K_1=1$.
As for 2), the number of modes in the state can easily be changed in the time domain. Here, we compare $N=8,16,32$.
Regarding 3), the measurement strategy, we have so far assumed that spatial mode A was measured (as illustrated in Fig.\ \ref{fig:KDvLinF}a), thereby transforming the nodes in mode B. However, alternative measurement schemes can be employed, considering that with a fast optical switch each spatio-temporal mode could be directed either to the variable quadrature measurement or to the output.
One should expect that increasing the number of circuit parameters, i.e.\ the length of $\vec\theta$ relative to $N$, should increase the expressibility. We explore that by comparing the previous ``linear'' measurement strategy with a new
``Knight's distance'' (KD) measurement strategy, where the output modes alternate between the two spatial modes following the Knight's jump pattern in chess, as shown in Fig.\ \ref{fig:KDvLinF}b. 
Finally, as for 4), the input squeezing $\vec{r}_\text{in}$ strongly affects both the accessible induced circuits $\mathcal{P}$ and the reference ensemble $\mathcal{H}$, given the method for constructing this outlined above. Furthermore, the induced (output) squeezing levels $\vec{r}_\text{eff}$ also depend on the measurement strategy: Additional measurements, as in the KD strategy, reduce the resulting output energy. To compare the two measurement strategies on even terms, we therefore choose the input squeezing levels such that the average photon number per mode, $\bar{n}_\text{pm}$, is identical. Further, all input modes are squeezed the same.
For each $N$, we compare three output energy levels, $\bar{n}_\text{pm}=0.1,0.3,0.5$, which is obtained by choosing different squeezing levels for the different measurement strategies. For instance, for $N=8$, the squeezing levels are chosen to be $r_\text{in}=0.457,0.736,0.900$ for the linear pattern and $r_\text{in}=0.764,1.036,1.196$ for the KD pattern to obtain these $\bar{n}_\text{pm}$ values.

We present the result of these simulations in Fig.~\ref{fig:KDvLinF}. For each configuration, we sampled 1000 circuits by sampling each homodyne phase angle $\theta_i$ uniformly between $-\pi/2$ and $\pi/2$. For the reference ensemble, we similarly sampled 1000 unitaries $W$ and combined  each of them with one of the induced squeezings $S_\text{sq}$ / $\vec{r}_\text{eff}$. All $10^6$ pairwise fidelities for each of the three terms in eq.~\eqref{expressibility} were then calculated, and the distributions for the two intra-set terms are plotted as histograms in panels (c-e): solid curves for the $\mathcal{P}$ distributions obtained with the linear pattern and dashed for the KD pattern, while the $\mathcal{H}$ distribution is filled. The deviations \eqref{expressibility} representing expressibility are presented in panels (f-h).
%
%
%
We see that, as the mean photon number per mode increases,
the fidelities accumulate closer to $0$. This effect is especially pronounced as $N$ increases. 
Looking at the deviation metric, there is not a definite trend, but in general, the deviations get lower as the squeezing/photon number increases.
Furthermore, as $N$ increases, the weights of tails of the distributions due to the different measurement strategies also get closer to that of the reference distributions. This is reflected in the higher order moments of the different measurement strategies. Lastly, the KD states show a significantly higher overlap with the references as compared to the linear states across all values of $N$ and mean per-mode photon numbers. 
This is also clearly reflected in the values of the deviations across the different values of $t$, which are considerably lower for the KD strategy states than for states from the linear strategies, showing that a more advanced measurement strategy can indeed be highly beneficial.
A more detailed argument for the reasons behind this is given in Supplementary Material, sec. 1 \cite{supp}.


\section{Experimental demonstration}


We now experimentally demonstrate the measurement-induced squeezed-light network using the setup in Fig.~\ref{fig:cluster}d.
We employ a configuration with two cascaded asymmetric fiber interferometers, generating a 2D graph state within two optical spatial modes (denoted A and B), with the first interferometer introducing a delay of $K_1=1$ clock cycle and the second interferometer a delay of $K_2=8$ clock cycles. 
We inject squeezed vacuum states generated by optical parametric oscillators (OPOs) into modes A and B with, respectively, 2.3 dB and 3.0 dB of squeezing in conjugate quadratures \cite{supp}, i.e.\ keeping $\vec r_\text{in}$ fixed.

The squeezed states are generated in the continuous wave regime, but the temporal structure introduced by the delay lines effectively defines time bins of duration 246.9 ns, corresponding to the $\sim$50 m length of the first fiber delay. Within these time bins, we define an anti-symmetric temporal mode that matches the OPO bandwidth but filters out noisy frequencies close to DC \cite{cluster19}.
We detect both modes A and B with homodyne detectors: Mode A is measured in a variable basis $\vec{\theta}$, while the resulting measurement-induced graph state in mode B is characterized by $x$ (position) and $p$ (momentum) quadrature measurements. The 2D cluster state, ideally, extends infinitely. However, to generate finite parametrized states, we ``chop'' the 2D cluster using $x$-quadrature measurements, which remove correlations.
Given the Gaussian nature of the state, full characterization is achieved by determining its multimode covariance matrix, which involves measuring all correlations for complete direct determination. To limit the number of measurements, 
we leverage the expected symmetry of the covariance matrix, thereby reducing the number of required measurements considerably (for details, see  \cite{supp}).

\begin{figure*}[]
\includegraphics[width=\textwidth]{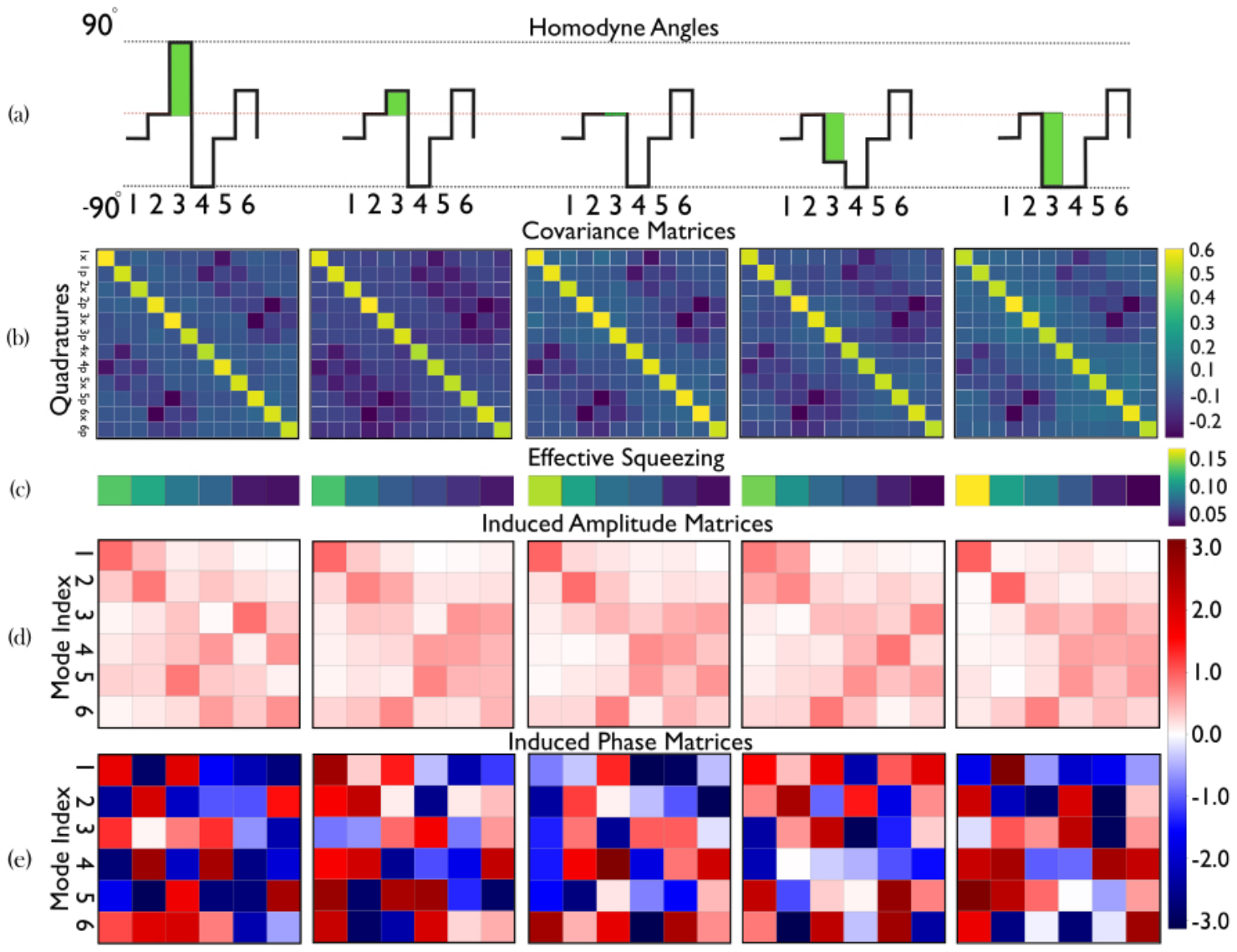}
\caption{\label{fig:states} Measurement induced states and corresponding interferometers:  five parametrized states and their induced unitaries generated as a result of varying measurement parameters on temporal mode 3 of spatial mode A on a 12 mode (6 temporal modes in A and 6 temporal modes in B) cluster state, demonstrating the ability of the method to induce parametrized interferometers. 
(a) The respective five different measurement settings applied to the 6 temporal modes. Note that the third temporal mode undergoes varying measurement bases, as highlighted in green. The covariance matrices of the parametrized states thus produced (b) are decomposed to form the unitary matrix whose amplitude (d) and phase elements (e) are shown indexed by the mode numbers of the output modes while the effective induced squeezing that enters the induced interferometers in also shown (c).}
\end{figure*}

We first implement a programmable six-mode squeezed light interferometer by measurements on spatial mode A of a 12-mode graph state, which was in turn obtained through deletions by $x$-quadrature measurements of neighbouring modes of the larger 2D cluster state \cite{GraphicalCalc}.
To illustrate the impact of different measurement basis choices, we carry out five different realizations in which a single homodyne angle, $\theta_3$, is varied in equal steps between $-\pi/2$ and $\pi/2$, as shown in Fig.~\ref{fig:states}a. For each of the realized output states, we present in Fig.~\ref{fig:states}b the covariance matrix with the thermalization from random measurement outcomes removed as described above and in \cite{supp}, sec. 3.6. The states are generated with relatively high fidelity of 0.91, 0.93, 0.89, 0.91 and 0.88, respectively, with respect to the simulated states with the input squeezing equal to the mean of squeezing values extracted from the experimental data (Supplementary Information, section 3.1.2). Most of the loss of fidelity can be attributed to loss and unequal squeezing of the two OPOs used for the generation of the cluster state.
Now, using Williamson and Bloch-Messiah-Euler decomposition of the covariance matrix, we construct the corresponding effective input squeezing and the unitary matrix of the induced interferometer, the amplitude and phase elements of which are illustrated in Fig.~\ref{fig:states}c and Fig.~\ref{fig:states}d for each of the five realizations. It is evident that altering just a single measurement angle has a significant impact on the configuration of the squeezed light interferometer, clearly demonstrating the ability to tune the quantum correlations of the resulting output.


\begin{figure*}

    \includegraphics[scale=0.69]{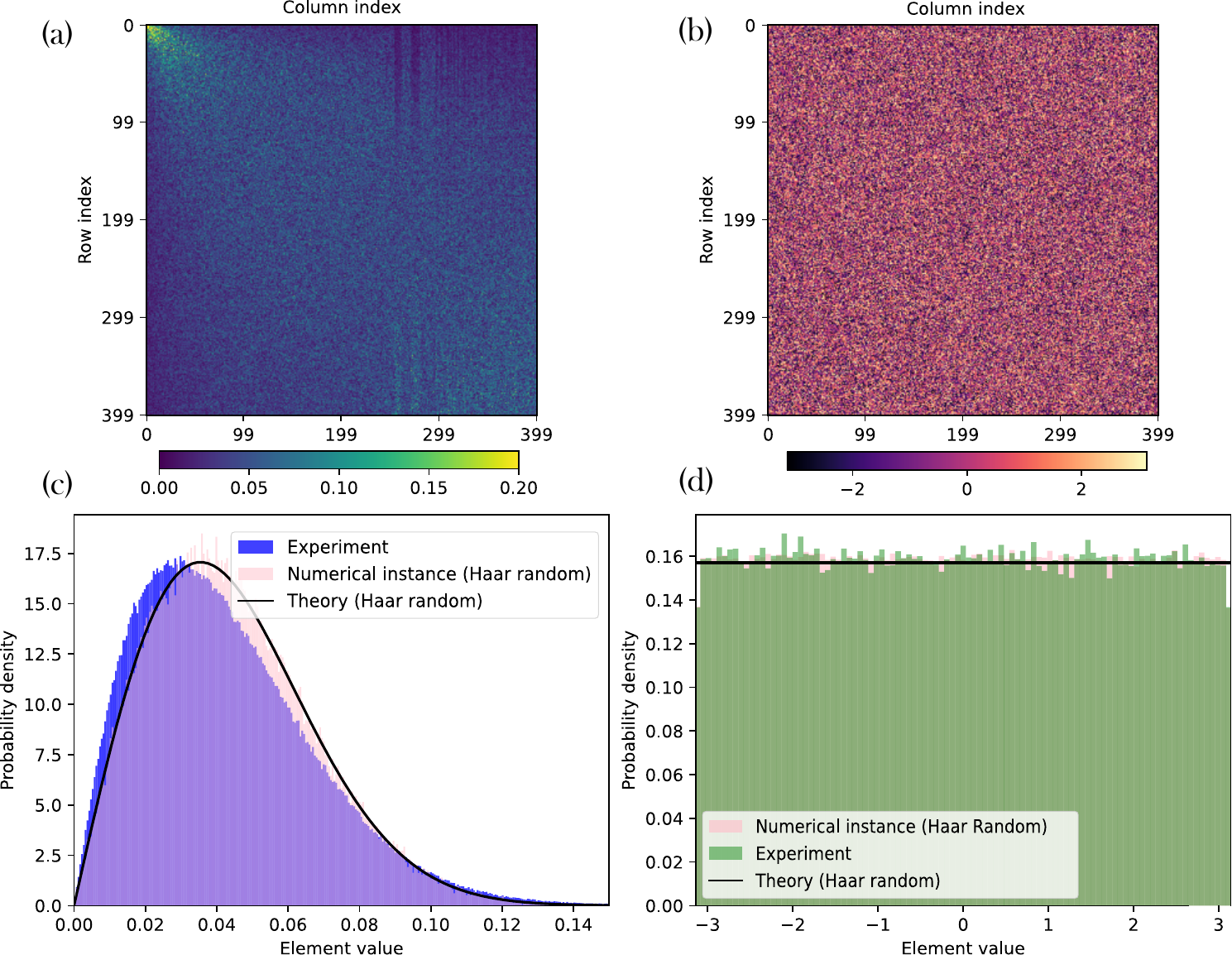}
     
     \caption{\label{fig:HaarDists} Elements of the random matrix : (a) Amplitude and, (b) phase parts of the unitary that was induced through these random measurements. (c) The probability distribution of amplitude and, (d) phase elements obtained in the experiment with respect to a instance of a numerically generated Haar random matrix and the theoretical prediction are shown here. We report fidelities of 0.997 and 0.999 in the cases of amplitudes and phases obtained, respectively, with respect to samples drawn from the Haar measure which is an improvement over the previous efforts. }
\end{figure*}

As a next step, we consider a 400-mode squeezed light interferometer, representing a realization that is practically difficult to achieve using a conventional bulk or photonic integrated interferometer. Similar to the six-mode realization, we prepare the output state through the measurements of mode A, but now using 400 different measurement angles, $\vec{\theta} = (\theta_1, \theta_2 , \ldots , \theta_{400})$, and reconstruct the covariance matrix through 22 measurement patterns across $x$ and $p$ quadratures (\cite{supp}, sec.\ 3.4). 
To implement a large random unitary matrix, the 400 angles are sampled uniformly from the interval $[-\pi/2, \pi/2]$, and the resulting covariance matrix of size $800\times800$ is characterized by the homodyne detector in mode B. Using this result, along with its decompositions, we deduce the distribution of the induced squeezed states, as well as the unitary matrix representing the beam splitter transformation. 
We first show the amplitudes and the phases of the random unitary generated in the experiment in Fig.\ \ref{fig:HaarDists}a and Fig.\ \ref{fig:HaarDists}b. 


In Fig.\ \ref{fig:HaarDists}, we also plot the amplitude and phase distributions of the elements of the resulting unitary matrix and compare them to the theoretical distribution for Haar-random unitaries as well as the distribution of numerically generated random matrices, which follow the distribution $P(a)=2(N-1)(1-a^{2})^{N-2}a$~\cite{USTC20,Xanadu21,LargeHaar} for the amplitude $a$ and a uniform distribution for the phase $\phi$. 

From Fig.\ \ref{fig:HaarDists}a, the unitary matrix exhibits a bit of structure, making it not completely random. To test the randomness, we show the distribution of eigenvalues and eigenphases of the matrices generated in supplementary information, section 3.7 (Figure 13, 14). We find the results consistent with that of a circular unitary ensemble. Furthermore, like previous efforts \cite{USTC20, Xanadu21} we use fidelity to quantify how closely the distribution of the experimentally produced matrix elements matches that of unitaries drawn according to the Haar measure. The fidelity is defined by $F=\sum_{i}\sqrt{p_{i}q_{i}}$, where $p_{i}$ and $q_{i}$ represents the probability densities of the element amplitudes and phases of the experimentally produced matrices and the matrices sampled from the Haar measure. The calculated fidelities are $0.997$ and $0.999$ for the amplitude and phase elements, respectively. Such matrices often find applications in demonstrations of quantum advantage using photons\cite{QAdv1,QAdv2,QAdv3,Cleo}. 

 \section{Conclusions}
 In this work, we have introduced and experimentally demonstrated a scalable, low-loss approach to generating multimode squeezed light interferometers based on measurement-induced operations. By leveraging quadrature measurements on time-domain graph states, we developed a flexible and programmable beam splitter network capable of accessing a wide class of unitaries. This method enables efficient parameter tuning through measurement, transforming a fixed, shallow circuit into a parametrized quantum circuit for generating parametric quantum states and interferometers. Notably, we demonstrated the realization of high-dimensional unitaries, including a random 400-mode interferometer with a distribution of elements that very closely resembles one whose samples are drawn from the Haar measure.

Our approach aligns with the paradigm of parametrized quantum circuits used in discrete-variable systems, where fixed circuit structures are parametrized and iteratively tuned to optimize specific tasks. Similarly, our method enables precise tuning of a fixed graph state circuit through measurement-based parameters. Although methods and metrics for generating parametric states have been extensively studied in discrete-variable systems \cite{PQC1,PQC2,PQC3,PQC4}, such an approach has been largely unexplored in continuous-variable systems.

The ability to efficiently generate and tune high-dimensional quantum states and interferometers has far-reaching potential in quantum computational tasks, including achieving quantum advantage~\cite{QAdv1,QAdv2,QAdv3}, simulating complex physical systems, and potentially addressing real-world challenges such as molecular modeling and drug discovery~\cite{Amb1,Amb2,Amb3,DD1,DD2}. This development demonstrates a scalable platform for continuous-variable quantum technologies, positioning parametrized measurement-based circuits as a potentially useful tool for advancing the capabilities of NISQ systems.

Although the current implementation is limited in producing exact states and interferometers for very large mode numbers, it demonstrates a pathway for further developments. By incorporating more advanced resource states and different initial graph states, refined measurement strategies and feedback schemes, as well as leveraging optimization techniques in order to produce states reasonably close to states of any particular interest, the range of achievable states and unitaries can be expanded, enhancing the practical applicability of the approach.

\section{Acknowledgements}
We acknowledge support from the Danish National Research Foundation (bigQ, DNRF0142), EU ERC project ClusterQ (grant agreement no. 101055224, ERC-2021-ADG), Innovation Fund Denmark (PhotoQ, 3155-00024A, and QuantERA - ClusSTAR, 3155-00024A), and EU (CLUSTEC, grant agreement no. 101080173, and QuantERA - ClusSTAR).


\bibliography{references}

\end{document}